\documentclass[10pt]{article}
\usepackage{amssymb}
\usepackage{amsmath}
\usepackage{amsfonts}
\usepackage[utf8]{inputenc} 
\usepackage{amsfonts}
\usepackage{graphtex}
\usepackage{pslatex}
\usepackage{subfigure}
\usepackage{graphicx}
\usepackage{listings}
\usepackage{color}
\usepackage{enumerate}
\usepackage[vlined,linesnumbered,algopart]{algorithm2e}

\usepackage{RRLif}
\newbox\subfigbox
\makeatletter
\newenvironment{subfloat}%
	{\def\caption##1{\gdef\subcapsave{\relax##1}}%
  		\let\subcapsave\@empty%
  		\setbox\subfigbox\hbox%
  		\bgroup%
	}%
	{%
		\egroup%
  	\subfigure[\subcapsave]{\box\subfigbox}%
  	}%

\makeatother

\definecolor{grey}{rgb}{0.95,0.95,0.95} % on définit la couleur grise (c'est un gris très clair)
\def\L{{\cal L}}
\def\OP{Op~}
\def\ID{ID}
\def\IT{IT}
\def\ChLab{ChLab}
\def\NoValue{NoValue}
\def\IW{{\cal W}}
\def\Cedit{FCedit}
\newcommand{\deps}{\succ_s}

\newcommand{\pars}{\parallel_s}
\def\union{\cup}
\def\Nat{{\cal N}}

\newcommand{\lmult}{\{}
\newcommand{\rmult}{\}}

\newcommand{\idp}{{id_p}}

\newtheorem{prop}{Proposition}%[section] 

\newenvironment{myproof}[1]{
\vspace{-5ex}
\hspace*{1em}\begin{trivlist}\item[\hskip\labelsep{\textsc{#1}}]}%
{\end{trivlist}}

\newenvironment{proof}{\begin{myproof}{Proof.}}{\hfill $\Box$\end{myproof}}

\newcommand{\idComp}[2]{(#1:#2)}

\newcommand{\MyFunction}[2]{{{\sc #1}$(#2)$:}}

\begin{document}

%###############################################################################
\begin{figure}
\centering
\end{figure}
\newpage

\title{Research report : Collaborative Peer 2 Peer Edition: Avoiding Conflicts is Better than
Solving Conflicts
}
\author{
S.~Martin and D.~Lugiez\\
LIF UMR 6166 Aix-Marseille Universit\'{e} CNRS}
\date{}
\maketitle
%\makeRR 
\thispagestyle{empty} 
\begin{abstract} 
Collaborative edition is achieved by distinct sites that work independently on
(a copy of) a shared document. Conflicts may arise during this process and
must be solved by the collaborative editor. In pure Peer to Peer collaborative
editing, no centralization nor locks nor time-stamps are used which make
conflict resolution difficult. We propose an algorithm which relies on the
notion or semantics dependence and avoids the need of any integration
transformation to solve conflicts. Furthermore, it doesn't use any history
file recording operations performed since starting the edition process.  We
show how to define editing operations for semi-structured documents
i.e. XML-like trees, that are enriched with informations derived for free from
the editing process. Then we define the semantics dependence relation required
by the algorithm and we present preliminary results obtained by a prototype
implementation.
\end{abstract}

%\keywords{Collaborative Editing, XML, Peer to Peer, Optimistic Reconciliation, Consistence Maintenance, Distributed Computing}
\section{Introduction}
\label{sec:intro}

Collaborative edition becomes more and more popular (writing article with SVN,
setting appointments with doodle, Wikipedia articles,\ldots) and it is
achieved by distinct sites that work independently on (a copy of) a shared
document. Several systems have been designed to achieved this task but most of
them use centralization and locks or weak centralization via time-stamps. A
alternative approach is the Peer to Peer approach -P2P in short- where new
sites can freely join the process and no central site is required to
coordinate the work. This solution is more secure and scalable since the lack
of central site prevents from failures and allows for a huge number of
participants. In this paper we focus on editing semi-structured documents,
called XML trees from now on, using the basic editing operations {\em add,
delete} for edges or {\em changing labels} in the document. Since the process
is concurrent, conflicts can occur: for instance a site $s_1$ changes the
label $Introduction$ of an edge by $Definition$ when another site $s_2$ want
to relabel $Introduction$ by $Abstract$. Then $s_1$ informs $s_2$ of the
operation performed and conversely. Executing the corresponding operations
leads to an incoherent state since the sites nor longer have identical copies
of the shared document. In the optimistic P2P approach, each operation is
accounting for and conflicts are solved by replacing the execution of an
operation $op_2$ performed concurrently with $op_1$ by $\IT(op_2,op_1)$ where
$\IT$ is an integration transformation defined on the set of operations.  This
transformation computes the effect of the execution of $op_1$ on $op_2$,
i.e. the {\em dependence of $op_2$ from $op_1$}.

In the word case, the transformations proposed in
\cite{ResselNG-CSCW96,EllisG-SIGMOD89,ImineMOR-CSCW03,DULIR-ICDCS04,SuleimanCF-GROUP97}
turned out to be non-convergent, see \cite{Imine-PhD06} for counter-examples. In
particular, none of these transformations satisfy both properties $TP1$ (a local
confluence property) and $TP2$ (integration stability)
that are sufficient to ensure convergence \cite{ResselNG-CSCW96}. Currently,
no convergent algorithm based on the integration transformation is known for
words. For XML trees, algorithms and operations have been proposed (like in
\cite{AgToXMLnotTP1}), but they have the same problem as in the word case or
use time-stamps (see \cite{OsterSMN-ICEIS07}) i.e. are not true P2P.

We propose a new algorithm that relies on {\em semantic dependence} of
operations which allows to reduce the integration transformation to a trivial
one: $\IT(op_2,op_1)=op_2$. This is possible since we enrich the data
structure by adding informations coming for free from the editing process on
trees yielding an important property: each edge is uniquely
labelled. Furthermore labels also record the level of dependence of the sites
that created or modified them. These properties allow to get a simple
convergent editing algorithm which doesn't require any history file recording
all operations done since the beginning of the edition process. Since a word
can be encoded as a tree, this algorithm also solves the word case, at the
price of a more complex representation. These ideas have been implemented in a
prototype that proved that the editing is done efficiently and that the
process is scalable.

Section \ref{sec:related} discusses the current approaches to collaborative
editing, and we present our editing algorithm in section
\ref{sec:conflict}. The data structure used for XML trees is described in
section \ref{sec:XML} and our first results are given in section
\ref{sec:experiments}. Missing proofs can be found in the full research report.

\section{Related Works}
\label{sec:related}

%\subsection{Other algorithms}
Many collaborative edition framework have been proposed, and we discuss only
the most prominent ones.

\paragraph{Document synchronization framework.} 

{\em IceCube} (see \cite{KermarrecARASMDP-PODC01}) is a operational-based
generic approach for reconciliating divergent copies. Conflicts
are solved on a selected site using optimization techniques relying on
semantic static constraints (generated by document rules) and dynamic
(generated by the current state of the document). Complexity is NP-hard and
this approach is not a true P2P solution (each conflict is solved by one
site). The {\em Harmony} project \cite{FosterGKPS-JCSS07} is a state-based generic
framework for merging two divergent copies of documents. These
documents are tree-like data structure similar to the unordered trees that we
discuss in section \ref{sec:XML}. The synchronization process exploits
XML-schema information and is proved terminating and convergent for two sites.

\paragraph{Integration transformation based framework.} 

{\em So6} \cite{OsterSMN-ICEIS07} is a generic framework based on the {\em
Soct4} algorithm which requires the local confluence property (TP1). It relies
on continuous global order information delivered by a times-tamper, which is
not pure P2P since it relies on a central server for delivering these
time-stamps.  The {\em Goto} system (Sun et al.\cite{SunC-CSCW98}), or {\em
SDT} (Du Li and Rui Li \cite{DuR-CSCW04}) rely on forward and
backward transformation (for undoing operations). These algorithms need to
reorder the history of operations which involve a lot of computations to
update the current state in order to ensure convergence.

%Goto Sun and all\cite{SunC-CSCW98} and Adopted, Ressel and all\cite{ResselNG-CSCW96} and and SDT\cite{DuR-CSCW04} Du Li \& Rui Li use local confluence ($TP1$) and Integration stability ($TP2$) to guarantee convergence.
%The problem is assure a same context when the operation integrating and when it has executed. To do that each algorithm has a different way to proceed.
%oto use forward and backward transformation(IT and ET) to re-order the history
%with three scenario. The backward transformation reverse forward
%transformation when it's possible.

%Adopted is multidimensional solution to calculate integration. Every user is a
%dimension, and the goal is integrate a new operation with find a path between
%last point (current state) and a final coordinate (all operations have
%executed) with only Forward transformation modeled as vector translation.
%They are a pure P2P algorithm, but they require a too hard properties for linear object as string  (TP1 \& TP2) to assure a convergence. The operation's set for text in groupware which have TP1 and TP2 properties is open problem.

{\em Goto} (Sun et al. \cite{SunC-CSCW98}), {\em Adopted} (Ressel et
al. \cite{ResselNG-CSCW96}) and {\em SDT} (Du Li and Rui Li \cite{DuR-CSCW04})
rely on the local confluence property ($TP1$) and on the integration stability
property ($TP2$) to guarantee convergence. A  main issue is to ensure that
operation integration takes place in the same context and return the same
result and each algorithm has its own solution. For instance, {\em Goto} uses a forward $(\IT$) and a
backward $ET$) transformation to reorder the history (record of all operations
performed). {\em Adopted} computes the sequence of integrations as a path in a
multi-dimensional cube. The main drawback of these approach is that it is hard
to design set of useful operations and integration transformations that satisfy both
$TP1$ and $TP2$. For instance, no such set exists in the word case nor for
linearly ordered structures.

The  set of operations given by Davis and Sun  provides
operations on trees for the Grove editor \cite{AgToXMLnotTP1}, but this set
doesn't satisfy the local confluence property TP1. Therefore, there is little hope
to get  a convergent editing process. {\em OpTree}
\cite{IgnatC-CEW02,IgnatN-CI06} present a framework for editing trees and
graphical documents using {\em Opt} or the {\em Soct2}, and relies extensively
on history files containing all operations performed on the date. The complexity
is at least quadratic in the size of the log file and no formal proof of
correctness is given.

A main problem of all these solutions -even when convergence is guaranteed- is
that they rely on manipulation of history files that records all operations
performed and these computations can become quite expensive.

\section{Conflict-free Solution}
\label{sec:conflict}

We propose a generic schema for collaborative editing which avoid the pitfalls
of previous works by avoiding the need to solve conflicts. First we give an
abstract presentation of this editing process and of the properties required
to ensure its correctness, then we show how it works for XML trees.

Each site participating to the editing process executes the same algorithm
(given in figure \ref{fig:Algo}) and performs operations on his copy of the
shared documents. Operations belong to a set of operations \OP, and we assume
that there is a partial order $\deps$ (i.e. an irreflexive, antisymmetric,
transitive relation) on operations and we write $op_1\pars op_2$ iff
$op_1\not\deps op_2$ and $op_2\not \deps op_1$. This ordering expresses causal
dependencies of the editing process: $op_1\deps op_2$ iff $op_2$ depends from
$op_1$ (for instance $op_1$ creates an edge and $op_2$ relabels this edge). In
our model the set $OpDep$ as $op\in Op, \forall op'\in OpDep | op\deps op'$ is
bounded set.  We show how to compute this relation for XML trees in section
\ref{subsec:dependence}.  A sequence of operations is denoted by
$[op_1;\ldots;op_n]$ and the result of applying $op_1$, followed by $op_2$,
\ldots, $op_n$ to the document $t$ is denoted by $[op_1;\ldots;op_n](t)$.
The set of operations $(\OP,\deps)$ is {\em independent} iff $\forall
op,op'\in\OP~\forall t,op \pars op' \implies [op,op'](t)=[op',op](t)$.

A sequence $[op_1;\ldots;op_n]$ is valid if for all $op_i,op_j$ occurring in
the sequence, $op_i\deps op_j$ implies $i<j$. In other words, the sequence is
{\em a linearization} of the partial order defined by $\deps$ on the set
$\{op_1,\ldots,op_n\}$. Given a valid sequence $[op_1;\ldots;op_n]$, a
substitution $\sigma$ of $\{1,\ldots,n\}$ is compliant with $\deps$ iff the
sequence $[op_{\sigma(1)};\ldots;op_{\sigma(n)}]$ is valid. This yields that
$op_i\pars op_j$ iff $op_{\sigma(i)}\pars op_{\sigma(j)}$ or in other terms,
$\sigma$ doesn't change the causality relation between operations. The
collaborative editing algorithm that we propose relies on the following
proposition\footnote{This result is a classical result in the field of
partial order}:

\begin{prop}
\label{prop:indep-inv}
Let $(\OP,\deps)$ an independent set of operations. Let $[op_1,\ldots,op_n]$
be a valid sequence of operations in $\OP$ and let $\sigma$ be a substitution
compliant with $\deps$. Then
$[op_1,\ldots,op_n](t)=[op_{\sigma(1)},\ldots,op_{\sigma(n)}](t)$
\end{prop}

\begin{proof}

Firstly, we prove that exchanging two consecutive non-dependent operations
doesn't change the result.

Let $\tau_i$ the substitution such that $\tau_i(i)=i+1, \tau_i(i+1)=i$ and
$\tau_i(k)=k$ otherwise. Let $[op_1;\ldots;op_n]$ be a valid sequence and let
$op_i\parallel op_{i+1}$.  We prove that
$[op_1;\ldots;op_n](t)=[op_{\tau_i(1)};\ldots;op_{\tau_i(n)}(t)]$ 
as follows:

$
\begin{array}[t]{ll}
[op_{\tau_i(1)};\ldots;op_{\tau_i(n)}](t)
&= [op_1;\ldots;op_{i-1};op_{i+1};op_i;op_{i+2}\ldots;op_n](t)\\
&=[op_{i+1};op_i;op_{i+2}\ldots,op_n](t') ~with~s'= [op_1;\ldots,op_{i-1}](t)\\
&=[op_{i};op_{i+1};op_{i+2}\ldots,op_n](t')~since~(\OP,\deps)~is~independent\\
&=[op_1,\ldots,op_n](t)\\
\end{array}
$\\

Secondly we prove the result by induction on the number of elements
in the sequence $[op_1;\ldots;op_n]$.

\begin{itemize}
\item Base case: $n=1$ straightforward.

\item Induction case: Let $[op_1;\ldots;op_n]$ be a valid sequence of $\OP$.

Let $[op_{\sigma(1)};\ldots;op_{\sigma(n)}]$ be another linearization of
$\{op_1,\ldots,op_n\}$.

We prove that
$[op_1;\ldots;op_n](t)=[op_{\sigma(1)};\ldots;op_{\sigma(n)}](t)$.

By definition $op_1$ is a maximal element of $\deps$. This element occurs at
position $j$ in $l=[op_{\sigma(1)};\ldots;op_{\sigma(n)}](t)$.  Let $\tau_k$
be the subtitution that exchanges the elements of $l$ at positions $k$ and
$k+1$ and leaves other elements unchanged.

Since $op_1$ is maximal, any operation $op'$ occurring in $l$ at position $k<j$
is such that $op'\parallel op$.

Therefore there is a sequence $\tau_{j-1},\ldots,\tau_1$ of substitutions such
that the application of these substitutions to
$[op_{\sigma(1)};\ldots;op_{\sigma(n)}]$ yields a sequence
$[op_1;op'_2;\ldots;op'_n]$ such that  (i) $[op_1;op'_2;\ldots;op'_n](t)=
[op_{\sigma(1)};\ldots;op_{\sigma(n)}](t)$ (by our first result) and (ii)
$[op_1;op'_2;\ldots;op'_n]$ is a linearization of $op_1,\ldots;op_n$.  

Therefore $[op'_2;\ldots;op'_n]$ is a linearization of $op_2,\ldots,op_n$. 

By induction hypothesis, we get
$[op'_2;\ldots;op'_n](t')=[op_2,\ldots,op_n](t')$.

Taking $s'=[op_1](t)$ yields the result.

\end{itemize}
\end{proof}

Another statement of the proposition is that the execution of any
linearization of a partial order on some initial value yields the same result.

\paragraph{The dependenceOf function.}

In our setting, operations are issued by sites and are numbered with an
operation number on this site. For instance, to delete a node in a tree, the
operation is defined by the action {\em delete}, the site identifier $SiteId$
of the site which issues this deletion and the operation number $OpCount$ on
this site. Furthermore, the data structure (the shared document) is build
using these operations and stores this information for each component (nodes
or edges for trees for instance). A request $r$ is a triple composed of an
operation {\em op}, a site identifier {\em SiteId}, and an operation number
{\em OpCount}. We assume that there is an function $dependenceOf(r)$ which
returns for each request $r$, the pair $(SiteId':OpCount')$ of any operation
$op'$ such that $op'\deps op$. Actually, this operation can return such pairs
only for the minimal (ofr $\deps$) operations $op'$ such that $op'\deps op$.
In section \ref{}, we show how to define effectively and in a simple way this
function for XML trees.

\paragraph{The (Fast Collaborative Editing) \Cedit Algorithm.}

The procedures (except {\em Main()}) of the generic distributed algorithm {\em
\Cedit} are given in figure
\ref{fig:Algo}. Each site has an unique identification stored in {\em SiteId}, a
operation numbering stored in {\em Opcount}, a copy of the document $t$ and a
list {\em WaitingList} of requests awaiting to be treated. The function {\em
dependenceOf(r)} with $r=(op, SiteId:OpCount)$ returns the pairs
$(nSite:cSite)$ with $nSite$ a site identifier, $cSite$ some operation count,
such that $op$ depends from an operation issued from site $nSite$ with
operation count $cSite$. This function is defined simultaneously with the data
structure, set of operations and dependence relation, see section
\ref{subsec:dependence} for the definition used for XML-trees.
The {\em Main()} procedure (not given in figure \ref{fig:Algo}) calls {\em
Initialize()} and enters a loop which terminates when the editing process
stops. In the loop, the algorithm choose non-deterministically to set the
variable {\em op} to some user's input and to execute {\em
GenerateRequest(op)} or to execute {\em Receive(r)}. {\em GenerateRequest(op)}
simply updates the local variables and broadcast the corresponding request to other
sites.  {\em Receive(r)} adds $r$ to {\em WaitingList} and executes all
operations of requests that becomes executable thanks to $r$ (relying on {\em
Execute} and {\em IsExecutable}).
\begin{figure}[htb]
	
	\fbox{
	\fontsize{8pt}{8pt}
		\begin{subfloat}
			\begin{minipage}{0.47\linewidth}
				\begin{algorithm}[H]
					\MyFunction{Initialize}{}\\
					\Begin{
				  		$\forall i,SReceived[i]=0$\tcp*[r]{State Vector of received operations}
				  		$(SiteId,Obj,OpCount,WaitingList)=(n,o,1,\{\})$
					}
				\end{algorithm}
				\begin{algorithm}[H]
					\MyFunction{GenerateRequest}{op}\tcp*[r]{User emit operation}
					\Begin{
				 		Let $r=(op,SiteId:OpCount)$\\
						\If{$isExecutable(r)$}{
					 		$OpCount = OpCount+1$\\
					 		$t=op(t)$\tcp*[r]{Apply operation}
					 		broadCast r to other participant.
						}
					}
				\end{algorithm}
				
				\begin{algorithm}[H] 
				\MyFunction{Receive}{r}\tcp*[r]{This function
is executed when  a request is received}
				\Begin{
					$WaitingList= WaitingList \cup {r}$\\
					\ForAll{$r \in WaitingList | isExecutable(r)$}{$execute(r).$\tcp*[r]{execute all executable request}}
				}
				\end{algorithm}
			\end{minipage}
		\end{subfloat}
		\begin{subfloat}
			\begin{minipage}{0.47\linewidth}
	
			\begin{algorithm}[H] 
				\MyFunction{isExecutable}{r}\tcp*[r]{Check that request r is executable}
				\Begin{
				 	 Let
$r=(op,\#Site:\#Op)$\\\tcp*[r]{Check that the previous operation on same site
has been executed}
				 	\If{$\#Site\neq SiteId \wedge SReceived[\#Site]\neq \#Op-1$}{
				 		\Return false}
					\tcp{Check all dependencies was executed}
				 	\For{$(nSite:cSite) \in dependancesOf(r)$}{
				 		\If{$SReceived[nSite] < cSite$}{
							\Return false}
					}
			 	\Return true
				}
			\end{algorithm}

			\begin{algorithm}[H] 
				\MyFunction{Execute}{r}\tcp*[r]{Execute a request r}
				\Begin{
				 	$r=(op,\#Site:\#Op)$\\
				 	$StateReceived[\#Site]=\#Op$\tcp*[r]{Update state vector}
				 	$WaitingList=WaitingList/r$\tcp*[r]{remove r from waiting list}
				 	$t=op(t)$ \tcp*[r]{Applies a operation}
				}
			\end{algorithm} 
		\end{minipage}
	\end{subfloat}
	}
	\caption{The Concurrent Editing Algorithm}
	\label{fig:Algo}
\end{figure}
%ede\\

The convergence property states that each site has the same copy $t$ of the
shared document after all operations have been received and executed by each
site. Firstly, we show that requests are executed in a sequence that respects
the dependence relation.

\begin{prop}
%\label{algo:compliant}
Let $op^s_1,\ldots,op^s_n$ be the sequence of operations generated by site $s$
using {\em GenerateRequest} . Then the operation count associated to $op_i^s$ is $i$
and $op^s_i \deps op^s_j$ implies $i<j$ .
\end{prop}
\begin{proof}
The first fact is obvious since {\em OpCount} is incremented by $1$ at each
creation of an executable request, starting from $0$. Line 6 to 9 of {\em
isExecutable(r=(op,\#Site,\#Op))} tests that each operation $op'$, issued by
site $nSite$ with operation number $cSite$, which is dependent of {\em op}
contained in $r$ has been executed. This is ensured by returning false if
$SReceived[nSite]< cSite$.
\end{proof}

\begin{prop}
\label{prop:compliant}
Let $s, s'$ be two distinct sites. Let $op^s_1,\ldots,op^s_n$ be the sequence
of operations generated by $s$ using {\em GenerateRequest}. Let
$op^{s'}_1,\ldots,op^{s'}_m$ be the sequence of operations executed by $s'$
using {\em GenerateRequest} or {\em Receive}. If $op^{s'}_{j_i}$ is the
execution of $op_i^s$ (from $s$) by $s'$ then the sequence
$op^{s'}_{j_1},\ldots,op^{s'}_{j_n}$ satisfies $j_1<j_2<\ldots <j_n$ (i.e. the
execution order on $s'$ respects the creation order on $s$, hence the
dependence relation).
\end{prop}

\begin{proof}

Before any execution of an operation (line 6 of {\em GenerateRequest} or line
5 of {\em Receive}) a call to {\em isExecutable} is performed. The first step
of this function returns false for an operation of site $s$ numbered $n$ if
the operation of site $s$ numbered $n-1$ has not been executed. Therefore the
execution order of the operations $op_i^s$ respects their creation
order. Since the creation order respects the dependence relation, we are done.
\end{proof}

\begin{prop} 
The algorithm {\em \Cedit} is convergent if the set of operations is independent.
\end{prop} 

\begin{proof}
Let $[op_1;\ldots;op_m]$ by the sequence executed on  site $s$.
We prove that $[op_1;\ldots;op_m]$ is a linearization of the partial order
defined by $\deps$ on $\{op_1,\ldots,op_m\}$.

Let $op_i$ and $op_j$ such that $op_i$ and $op_j$ have been generated by the
same site $s'$.  The subsequence $[op_{j_1};\ldots;op_{j_l}]$ corresponding to
the operations received from site $s'$ is such that $op_{j_k}\deps
op_{j_{k'}}$ implies $j_k<j_{k'}$ (by proposition \ref{prop:compliant}).

Let $op_i$ and $op_j$ such that $op_i$ has been generated by $s'$ and $op_j$
has been generated by $s''$. If $op_i\deps op_j$, the function {\em
isExecutable} called on the request $r=(op_j,\ldots)$ before executing $r$ on
site $s$ checks that $op_i$ has been executed on site $s$ (line 6 to 9 of {\em
isExecutable}). Therefore we get that $i<j$.

Therefore $[op_1;\ldots;op_m]$ is a linearization of the partial order induced
by $\deps$ on $\{op_1,\ldots,op_m\}$.  Since each site executes a
linearization of the same partial order, proposition \ref{prop:indep-inv}
yields that each site computes the same value for the shared document.

\end{proof}

\section{Conflict free operations for XML Trees}
\label{sec:XML}

The basics editing operations on trees are insertion, deletion or relabeling
of a node. Actually, since we consider edge labelled trees instead of node
labelled trees, insertion and deletion are performed on edges instead of
nodes. Firstly, we consider unordered trees, and we show in section
\ref{subsec:order} how to reestablish the ordering between edges, which allows to get
a data-structure corresponding to XML trees.

\subsection{Data Structure}

The information stored in nodes (or edges in our case) can be described as a
word on some finite alphabet $\Sigma$. To get a independent set of operations
containing relabeling, we must have a much more complex labeling 
that we describe now.

{\bf The set of identifiers $\ID$.} Each site is uniquely designated by its
identifier which is a natural number (IP numbers could be used as well). The
set of identifier is the set $\ID$ of pairs $(\idComp{SiteNumber}{NbOpns})$
where $NbOpns\in Nat$ is denotes some numbering of operations on this site.

{\bf The set of labels $\L$.} A label is a pair $(l,id)$ where $id\in\ID$ and
$l$ is a triple $(lab,id',dep)$ with $lab \in \Sigma_L^*$ with $\Sigma_L$ a finite
alphabet, $id'\in \ID$, $dep\in\Nat$ (expressing a level of dependence).

{\bf Trees.}
Trees are defined by the grammar 
\[
T\ni t::=\{~\}~|~\{n_1(t_1),\ldots,n_p(t_p)\}~~where~n_i=(l_i,id_i)\in \L, t_i\in T
\]
where  {\bf each $id_i$ occurs once in $t$}. 

The uniqueness of labels is guaranteed by the fact that
$id_i=(\idComp{SiteNumber}{NbOpns})$ states that the edge has been created
by operation $NbOpns$ of site $SiteNumber$.

Trees are unordered i.e. $\{n_1(t_1),\ldots,n_p(t_p)\}$ is identified with
$\{n_{\sigma(1)}(t_{\sigma(1)}),\ldots,n_{\sigma(p)}(t_{\sigma(p)})\}$ for any
permutation of $\{1,\ldots,n\}$.

{\bf Example.} We give an XML document and a tree that may represent this document as the
result of some editing process.

\begin{figure}[htb]
	\fontsize{9pt}{9pt}
		\begin{subfloat}
			\begin{minipage}{0.5\linewidth}
				\fontsize{8pt}{8pt}
				\begin{lstlisting}
<?xml version="1.0" encoding="UTF-8"?>
<Pat>
	<Phone>
		<Cellular>
			0691543545
		</Cellular>
		<Home>
			0491543545
		</Home>
	</Phone>
</Pat>
<Henri>
	<Adress>
		45 Emile Caplant Street
	</Adress>
</Henri>
				\end{lstlisting}
			\end{minipage}
			\caption{XML Document}
		\end{subfloat}
		\begin{subfloat}
			\begin{minipage}{0.5\linewidth}
				$\pic
				\Edge  (0,0) (-20,-20) (0,0) (20,-20) (-20,-20) (-20,-40) (20,-20) (20,-40) (-20,-40) (-30,-60) (-20,-40) (-10,-60) (20,-40) (20,-60) (-30,-60) (-30,-80) (-10,-60) (-10,-80)
				\Vertex(0,0) (-20,-20) (20,-20) (20,-40) (-20,-40) (-30,-60) (-10,-60) (20,-60) (-30,-80) (-10,-80)
				\Align[c] ($t=$) (0,10)

				\Align[r] (Pat) (-15,-7)
				\Align[l] (Henri) (15,-7)
				\Align[r] (Phone) (-25,-30)
				\Align[l] (Home) (-11,-47)
				\Align[r] (Cellular) (-30,-47)
				\Align[l] (0491543545) (-5,-70)
				\Align[r] (0691543545) (-35,-70)
				\Align[l] (Address) (25,-30)
				\Align[l] (45 Emile Caplant Street) (25,-50)
				\cip$
			\end{minipage}
		\caption{Schematic tree}
	\end{subfloat}
	\caption{Document}
\end{figure}
\begin{footnotesize}
$$t= 
\left \lmult \begin{array}{l} 
((Pat,\idComp{1}{3},2),\idComp{1}{1}) 
 \left ( \left \lmult \begin{array}{l} 
                      ((Phone,\idComp{3}{4},5),\idComp{2}{1}) \left ( \left \lmult 
                      			\begin{array}{l} 
                                                        ((Home,\idComp{3}{2},1)\idComp{3}{1})(\lmult((0491543545,\idComp{4}{2},1),\idComp{4}{1})(\lmult\rmult)\rmult) \\
                                                        ((Cellular,\idComp{5}{2},3),\idComp{5}{1})(\lmult((0691543545,\idComp{6}{2},1),\idComp{6}{1})(\lmult\rmult)\rmult)                               					\end{array}
                                           \right \rmult  
                                   \right )
                       \end{array} 
          \right \rmult 
   \right )\\
((Henri,\idComp{2}{3},1),\idComp{2}{2})(\lmult ((Address,\idComp{3}{5},2),\idComp{3}{2})(\lmult ((45~ Emile~ Caplant~ Street,\idComp{4}{9},5),\idComp{4}{2})
(\lmult\rmult)\rmult) \rmult) 
            \end{array}
\right \rmult
$$
\end{footnotesize}
\subsection{Editing Operations}

We extend the set $\Sigma_L$ by a symbol $\NoValue$ that states that a label is
not yet set.

{\bf Adding an edge.} The operation $Add(\idp,id)$ with $\idp\neq id$ adds
an edge labelled by $(l,id)$  with $l=(\NoValue,id,0)$ under edge labelled
$(\ldots,\idp)$. When $\idp$ doesn't occur, the tree is not modified. It is formally defined by:

\begin{array}[t]{l}
Add(\idp,id)(\{~\})=\{~\}
\\
Add(\idp,id)(\{n_1(t_1),\ldots,(l_i,id_i)(t_i),\ldots,n_p(t_p)\})
=
\{n_1(t_1),
\ldots,
(l_i,id_i)(t_i\union ((\NoValue,id,0),id)(\{~\})
\ldots
n_p(t_p)
\}\\
\hspace{4cm}~if~\idp=id_i\\
Add(\idp,id)(\{n_1(t_1),\ldots,n_p(t_p)\}) =
\{n_1(Add(\idp,id)(t_1)),
\ldots,
n_p(Add(\idp,id)(t_p))
\}\\

\hspace{4cm}~if~n_i=(l_i,id_i)~with~ id_i\neq \idp~for ~i=1,\ldots,n
\\
\end{array}

{\bf Deleting a subtree.}  The operation $Del(id)$ deletes the whole subtree
corresponding to the unique edge labelled by $(\ldots,id)$ (including this
edge).  When $id$ doesn't occur, the tree is not modified. It is formally
defined by:

$
\begin{array}[t]{l}
Del(id)(\{~\})=\{~\}
\\
Del(id)(\{n_1(t_1),\ldots,(l_i,id_i)(t_i),\ldots,n_p(t_p)\})
=
\{n_1(t_1),
\ldots,
n_{i-1}(t_i),n_{i+1}(t_{i+1}),
\ldots
n_p(t_p)
\}\\
\hspace{6cm}~if~id= id_i\\
Del(id)(\{n_1(t_1),\ldots,n_p(t_p)\}) =
\{n_1(Del(id)(t_1)),
\ldots,
n_p(Del(id)(t_p))
\}\\
\hspace{5cm}~if~n_i=(l_i,id_i)~with~ id_i\neq id~for ~i=1,\ldots,n
\\
\end{array}
$

{\bf Changing a label.} $\ChLab(id_{e},id_{op},dep,L)$ with $id_e,id_{op}\in 
\ID, dep\in \Nat, L\in \Sigma_L$ replaces the label 
$(l_e,id_e)$ of the edge identified by $(\ldots,id_e)$ by $(L,id_{op},v)$ depending
on some relations on dependencies. It is defined formally  by:\\ 

$
\begin{array}[t]{l}
\ChLab(id_{e},id_{op},dep,L)(\lmult n_1(t_1),
                                       \ldots
                                      (l_{e},id_{e})(t_{e}),
                                       \ldots
                                       n_p(t_p)\rmult)) = 
\lmult n_1(t_1),...(l'_{e},id_{e})(t_{e}),\ldots,n_p(t_p)
\\ 
$ where $
l_e=(L_{e},id_{e},dep_e)$ and $
l'_e=\left\{
	\begin{array}{l}
		(L,id_{op},dep)$, if $dep_e>dep$ or else $dep=dep_e$ and $ id_{op}< id_{lbl}\\
		l_{e} $, otherwise$
	\end{array}
\right. \\
\end{array}
\\
\begin{array}[t]{l}
\ChLab(id_{e},id_{op},dep,L)(\lmult n_1(t_1),...,n_p(t_p)\rmult))
=
(\lmult n_1(\ChLab(id_{e},id_{op},dep,L)(t_1))
         \ldots
         n_p(\ChLab(id_{e},id_{op},dep,L)(t_p))\rmult) \\
$ if $n_i=(l_i,id_i)$ with  $id_i\neq id_e$ for $i=1,\ldots,p\\
\end{array}
$

%{\bf Changing a label.} 

%%\subsubsection{ChLabel}

%$$ChLabel: ID \times ID \times N \times \Sigma^* \times T \longmapsto T$$
%%$$ChLabel(id_edge,id_op,nivDep,Lbl)$$
%$ChLabel(id_{edge},id_{op},nivDep,Lbl)$ Change label of $id_edge$ with  $id_{op}$ as operation id and depLvl as count of change seen by site to label Lbl. \\

%$ChLabel(id_{edge},id_{op},nivDep,Lbl)(\lmult(l_1,id_1)(t_1),...,(l_q,id_q)(t_q)\rmult))\\
%=(\lmult(l_1,id_1)(ChLabel(id_{edge},id_{op},nivDep,Lbl)(t_1)),...\\,(l_q,id_q)(ChLabel(id_{edge},id_{op},nivDep,Lbl)(t_q))\rmult)$ \\si $id_i\neq id_{edge}$ for $i\in [1..q]$\\

%$ChLabel(id_{edge},id_{op},nivDep,Lbl)(\lmult(l_1,id_1)(t_1),...(l_{edge},id_{edge})(t_{edge}),...,(l_q,id_q)(t_q)\rmult))$\\$=\lmult(l_1,id_1)(t_1),...(l'_{edge},id_{edge})(t_{edge}),...,(l_q,id_q)(t_q) $\\ $(lbl_{edge},id_{edge},niv_{edge})=l_{edge} $ \\

%With $l'_{edge}=
%\left\{
%	\begin{array}{l}
%		(lbl,id_{op},nivDep)$, if $niv>nivDep$ or $nivivDep\wedge id_{op}< id_{lbl}\\
%		l_{edge} $, other$
%	\end{array}
%\right.$\\

\subsection{Semantic Dependence}
\label{subsec:dependence}

Let the  set of operations be $\OP=\{Add(id,id'), Del(id),
\ChLab(id,id',dep,L)~|id,id'\in \ID,dep\in \Nat,L\in\Sigma_L^*\}$. The
dependence relation $\deps$ is defined as follows:  

\begin{itemize}

\item $ Add(id,\idp)\deps Del(id)$: an edge can be deleted only if it has been
created.

\item $ Add(\idp,\idp')\deps Add(id,\idp)$: adding  edge $id$ under edge $\idp$
requires that edge $\idp$ has been created.

\item $ Add(id,\idp)\deps \ChLab(id,id_{op},dep,L)$: changing the labeling of
edge $id$ requires that edge $id$ has been created.
\end{itemize}
 
This allows to compute the set of identifiers depending from an operation:

$dependencesOf(op) = 
	\left\{
		\begin{array}{l} 
			 \idp$ for $op=Add(\idp,id)\\
		 	id$ for $op=Del(id) \\
			id$ for $op=\ChLab(id,id_{op},depLvl,lbl)
		\end{array}
	\right.$

%---------propriété si op_1-/-> op_2 et op_2 -/-> op_1 => [op_1,op_2](s)=[op_2,op_1](s)             (*)

\begin{prop}
The set $(\OP,\deps)$ is an independent set of operations.
\end{prop}

\begin{proof}
We prove that if $op_1 \pars op_2$ then $ [op_1,op_2](t)=[op_2,op_1](t)$
by a case analysis on all possible pairs $op_1,op_2$. 

\begin{enumerate}
	\item $op_1=Add(id_1,\idp_1)$
		\begin{enumerate}
			\item $op_2=Add(id_2,\idp_2)$
				\begin{itemize}
				 	\item $\idp_1=id_2$ or $\idp_2=id_1$ there for respectively $op_1 \deps op_2$ or $op_2 \deps op_1$.
					\item else we can insert a edge before another independently of order the result will be same as a set.
				\end{itemize}

			\item $op_2=Del(id_2)$
				\label{AddDelTP1}
				\begin {itemize}
					\item $id_2=\idp_1$ or $\idp_1$ is in subtree $id_2$: \\
						let t a tree.
						$t_1=Del(id_2)(t)$ by definition $\idp$ is deleted.\\ $Add(id_1,\idp_1)=t_1$.
						$t_2=Add(id_2,\idp_1)(t)$ and $Del(id_2)(t_2)=t_1$ because a subtree are erased.
					\item $id_2=id_1$: 
						because $Add(id_1,\idp_1) \deps del(id_1)$.	
					\item other : 
						the edge $id_1$ has been created and $id_2$ has been deleted whatever order.
				\end{itemize}
			\item $op_2=ChLabel(id_2,id_{op2},dep_2,lbl_2)$
				\label{AddChlabelTP1}
				\begin{itemize}
					\item $id_2=id_1$ : the edge be created before renamed because $Add(id_1,\idp_1) \deps ChLabel(id_1,id_{op2},dep_2,lbl_2)$.	
					\item other, the add have no effect on ChLabel and vice versa.\hfill$\diamond$
				\end{itemize}
		\end{enumerate}
	\item $op_1=Del(id_1)$
		\begin {enumerate}
			\item $op_2=Add(id_2,\idp_2)$ : It's \ref{AddDelTP1} case.
			\item $op_2=Del(id_2)$ If $id_1$  is a subtree $id_2$ then $[del(id_1),del(id_2)](t)$ there are no edge to delete with $del(id_1)$ because it was deleted with $del(id_2)$ . And $[del(id_2),del(id_1)](t)$ the the edge and subedge of $id_1$ were deleted at first time and $id_2$ with $id_1$ was deleted too.
			else two subtree are distinct .
			\item $op_2=ChLabel(id_2,id_{op2},dep_2,lbl_2)$\\
				\label{DelChlabelTP1}
				\begin{itemize} 
					\item $id_1=id_2$\\
					Let $t'=del(id_1)(t)$.
					$Chlabel(id_1,id_{op2},dep_2,lbl_2)(t')=t'$ because $id_1$ is not present in $t'$.\\
					$del(id_1)(Chlabel(id_1,id_{op2},dep_2,lbl_2)(t))=t'$ because $id_1$ and it subtree was deleted. Whatever her label. 
					\item Other : there are no problems.
				\end{itemize}\hfill$\diamond$
		\end{enumerate}
	\item $op_1=Chlabel(id_1,id_{op_1},dep_1,lbl_1)$
		\begin{enumerate}
			\item $op_2=Add(id_2,\idp_2)$ : It's \ref{AddChlabelTP1} case.
			\item $op_2=Del(id_2)$ :  It's \ref{DelChlabelTP1} case.
			\item $op_2=ChLabel(id_2,id_{op2},dep_2,lbl_2)$ :
				\begin{itemize}
					\item $id_1\neq id_2$: The edge be different.
					\item $id_1=id_2$
						\begin{itemize}
							\item $dep_1<dep_2$
								let $t_1=op_1(op_2(t))^{(1)}$\\
								let $t_2=op_2(op_1(t))^{(2)}$\\
								In $^{(1)}$ the label of $id_1$ is $lbl_2$ and not changed by $op_1$ (definition).
								in $^{(2)}$ the label of $id_1$ is $lbl_1$ and changed by $op_2$ to $lbl_2$ (definition).\\
								therefore $t_1=t_2$.
							\item $dep_2<dep_1$: idem with number of label are inverted.
							\item $dep_1=dep_2$ 
							if $id_{op_1}<id_{op2}$
								same of $dep_1<dep_2$\\
							else
								same of  $dep_2<dep_1$ \\
							By definition $id_{op_1}\neq id_{op2}$\hfill$\diamond$
						\end{itemize}
				\end{itemize}
		\end{enumerate}
\end{enumerate}
\end{proof}

%Partie difficile pour en arriver la.

%########################################################################
\subsection{Ordered Trees}
\label{subsec:order}

The previous editing process is defined on unordered trees when XML documents
are ordered trees. To make the algorithm work in this case, we enrich the
labeling of edges with an ordering information. This shows that our approach
works in this general case. The properties required on  the ordering information are:

\begin{itemize}

\item The ordering of labels must be a total order

\item The ordering is the same for each site

\item Insertion can be done between two consecutive edges, before the smallest
edge and after the largest edge.

\end{itemize}
The ordering that we design enjoys all these properties. To each edge
corresponding to some identifier $id$ we associate a  word on some
finite alphabet $\Sigma$ such that two distinct edges corresponds to distinct
words.

Let $\Sigma_0=\{a_1,\ldots,a_n\}$ a finite alphabet such that there is a
injective mapping $\phi$ from $\ID$ into $\Sigma_0^*$. For instance, to a pair
$(\idComp{s}{n})$ with $s$ a site number, $n$ an operation number, we can associate a
word $dec(s)\cdot dec(n)$ on the alphabet $\{0,1,\ldots,9\}\union\{\cdot\}$
with $dec(x)$ the representation of $x$ in base $10$.

We extend $\Sigma_0$ by the letter $\#$ used as a separator and $\bot$ used as
a minimal element, yielding a alphabet $\Sigma$. The ordering on letters is
$\bot\le \#\le a_1\ldots <a_n$. The lexicographic ordering on words of
$\Sigma^*$ induced by the ordering of letters is a total ordering.

The labeling of an edge $e$ corresponding to the identifier $id_e$ is enriched
by a new field $p_e\in (\Sigma_0\union \{\bot;\#\})^*$ and we associate to $e$
the word $w_e=p_e\#\phi(id_e)$. The $\#\phi(id_e)$ part is added to guarantee
that distinct edges are associated to distinct words.

\begin{prop}
The ordering on edges defined by $e\prec e'$ iff
$w_e=p_e\#\phi(id_e)\ll w_f=p_f\#\phi(id_f)$ is a
total ordering on edges.
\end{prop}

\begin{proof}
Since distinct edges have distinct identifier, the function $\phi$ is
injective and $\#\phi(id_e)$ is the smallest suffix of $w_e$ containing only
one occurrence of $\#$, then the words associated to distinct edges are
distinct. This proves the proposition since $\ll$ is a total ordering on
words.
\end{proof}

{\bf Example.}
Let $e,f$ be edges identified by  $id_e=(1,10)$ and $id_f=(2,1)$. Let
$\phi(id_e)=1.10$ and $\phi(id_f)=2.1$. Let the priority of $e$ be $12$ and
the priority of $f$ be $211$. The ordering on digit is $'i'<'j'$ if $i <j$ and
$.<'i'$. Since $11\#1.10\ll 211\#2.1$, we get that edge $e$ precedes edge $f$
in the tree.

Let $\IW$ be the set of  words of the form $w_p\#w_{id}$
with $w_p\in \Sigma^*$, $w_{id}\in \phi(\ID)\subseteq \Sigma_0^*$.

\begin{prop}
Let $w,w'\in \IW$ such that $w\ll w'$.

\begin{enumerate}[(i)]
\item There exists a computable $w''\in \IW$ such that $w\ll w''$ and
$w''\ll w'$.

\item There exists $w_m,w_M\in \IW$ such that $w_m\ll w$ and $w'\ll w_M$.
\end{enumerate}
\end{prop}

\begin{proof}
Let $s[k]$ denote the $k^{th}$ letter of a word $s$ and let $|s|$ denote the
length of the word $s$.
\begin{enumerate}[(i)]
	\item Let $w=w_p\#w_i\ll w'=w_{p'}\#w'_i$. We construct $w''$ such that $w\ll
	w''\ll w'$.   Let $j$ be the minimal integer such that
	$w[j]<w'[j]$.
	\begin{enumerate}[{Case }1.]
	\item $j<length(w'_p\#w'_i)$.  Let $w_{p''}$ such that $|w''_p|=|w'_p\#w'_i|$ and
		$w''_{p}[k]=w'_p[k]$ for $k=1,\ldots,j$ and $w''_p[k]=\bot$ for $j<k\le
		length(w''_p)$. Given any $w''_i=\phi(id)$ for some $id$, by construction
		the  word $w''=w''_p\#w''_i$ is such that $w\ll w''<w'$.

	\item $j=length(w'_p\#w'_i)$.  Let $w_{p''}=w_p\#w_i\#$.  Given any
		$w''_i=\phi(id)$ for some $id$, by construction the word
		$w''=w''_p\#w''_i$ is such that $w\ll w''<w'$.

	\end{enumerate}
	\item Let $w=w_p\#w_i<w'=w_{p'}\#w'_i$. We construct $w_m$ such that
	$w_m\ll w$. Let $w_p^m[k]=\bot$ for $i=1,\ldots,length(w_p)+1$.  Given any
	$w^m_i=\phi(id)$ for some $id$, by construction the  word
	$w_m=w_p^m\#w^m_i$ is such that $w_m<w$.
	The same construction works to get $w_M$ such that $w'\ll w_M$ (use $a_n$
	instead of $\bot$).
\end{enumerate}
\end{proof}

\paragraph{An updated set of operations.} 
The data structure is slightly modified since the labels are now elements
$(l,id)$ with $id\in \ID$ and $l$ a tuple $(lab,id',dep,p)\in \Sigma_L^*,
id'\in \ID, dep\in \Nat, p\in \IW$. The field $p$  combined with the
identifier $id$ is used to order the edges arising from the same node,
therefore the data structure is similar to semi-structured documents.

The $Add$ and $\ChLab$ operations must be slightly modified to handle the new
field $p$, which simply amounts to considering a different set of labels. The
set of dependence between operation is the same as before and we have:

%The second clause defining  $Add$ is
%updated by replacing the tree
%$((\NoValue,id,0),id)(\{~\})$  by
%$((\NoValue,id,0,\bot),id)(\{~\})$.
%The relabelling operation $\ChLab(id_{e},id_{op},dep,L,p)$ with  $id_e,id_{op}\in 
%\ID, dep\in \Nat, L\in \Sigma_L, p\in \Sigma$ behaves like the original one,
%but also set the priority of the modified edge to $p$. 

\begin{prop}
The set $(\OP,\deps)$ is an independent set of operations.
\end{prop}

Therefore our collaborative editing algorithms works for ordered trees, i.e. XML
trees.

\section{Experiment and Future Works}
\label{sec:experiments}
%\TODO{Take a screenshot and state}
%\begin{figure}[t]
%	\fontsize{9pt}{9pt}
%		\begin{subfloat}
%			\begin{minipage}{0.5\linewidth}
% 			
% <<<<<<< Main.tex
% 					%\includegraphics[height=3in]{screen.pdf}
% 			\end{minipage}
% 				\caption{ScreenShot}
% 			\end{subfloat}
% 			\begin{subfloat}
% 				\begin{minipage}{0.5\linewidth}
% 			    \end{minipage}
% 				\caption{Statistic}
% 			\end{subfloat}
% 	\end{figure}
% We have implemented this algorithm and this operation's set on a prototype. 
% It's tested with a P2P simulator. 
% The P2P simulator shuffle $n$ messages when a broadcast to another site and verify all trees have equivalent after $m$ operations. The benchmarking \ref{benchmark} show the user time not depend of number of operation or number of site (compute message time).
% 
% We have implemented network layer. When a user connect to network, they exchange a state.
% 
% {\bf Future works:}We examine to add a smart log policies which be able to undo operation. We devise to resolve schema case without conflict with all site or reject locally alway the same operation. We will interest to put a security policy.
% =======
%					\includegraphics[height=3in]{screen.pdf}
%			\end{minipage}
%				\caption{ScreenShot}
%			\end{subfloat}
%			\begin{subfloat}
%				\begin{minipage}{0.5\linewidth}
%			    \end{minipage}
%				\caption{Statistic}
%			\end{subfloat}
%	\end{figure}

We have implemented the algorithm and the data structure for XML trees in java
(including the ordering information) on a Mac with a 2.53GHz processor.

The data structure {\em tree} is composed of {\em  edges}. Each edge have the following fields : 
\begin{itemize}
\item a field for storing its identifier (which is unique).
\item a field for storing the sons (which are edges).
\item a field for storing its ancestor (which is an edge).
\end{itemize}

A tree is identified as a some edge (the root). Access to an edge having some
identifier is done using a hash-table with identifier as key.  The initial
document is composed by only one edge: the root with like identifier $0:0$.
Applying an operation op on the tree is performed by the function {\em do} :
$Tree\times Op\longmapsto Tree$.

The implement of  {\em do} is straightforward. For instance $do(Add(id_f,id),tree)$: 
\begin{enumerate}[(i)]
\item creates a new edge with identifier $id$.
\item asks the hash-table to get the father edge $id_f$
\item stores the father reference.
\item adds new edge into the father list.
\item adds new edge references in the hash-table.
\end{enumerate}
%Our tree is composed by an edges. All edges know its father and have vector of sons. 
%In our model all edges is identified by a site number and operation number (identifier). The first edge without parent is called a root, its identifier is $0:0$. For increase the speed we use an hash-table have direct access on Edge from identifier.

% 
%%
%%The direct access to edge is powered by a hash-table, having the identifier for key.
%%All class operation can generate the dependance set.

%For example, the $dependenceof$ function with operation $Add$ return a set with the father identifier as dependent set and  $do : Tree \times Op \mapsto Tree$ function with this operation, create a new edge, ask the hash-table to get a father edge, set a new edge father with father references and add a new edge in father's sons list. 

%The other operations work with same schema. (for more informations about this, read the prototype documentation)

The P2P framework is simulated by random shuffling of the messages that are
broadcast. The results obtained with our prototype are given in Figure~\ref{fig:measures}.

%this operation's set on a prototype. 
%It's tested with a P2P simulator. 
%The P2P simulator shuffle $n$ messages when a broadcast to another site and verify all trees have equivalent after $m$ operations. The benchmarking \ref{benchmark} show the user time not depend of number of operation or number of site (compute message time).

%We have implemented network layer. When a user connect to network, they exchange a state.

\begin{figure}[thb]
\centering
	\fontsize{9pt}{9pt}
		\begin{subfloat}
			\begin{minipage}{0.3\linewidth}
				$$\pic

					\xaxis[0,100] 
					\yaxis[0,60] 
					\Line (10,-2) (10,2) 
					\Line (20,-2) (20,2) 
					\Line (30,-2) (30,2) 
					\Line (40,-2) (40,2) 
					\Line (50,-2) (50,2) 
					\Line (60,-2) (60,2) 
					\Line (70,-2) (70,2) 
					\Line (80,-2) (80,2)
					\Line (-2,10) (2,10) 
					\Line (-2,20) (2,20) 
					\Line (-2,30) (2,30) 
					\Line (-2,40) (2,40) 
					
					\Align[r] (0) (-5,-10)
					\Align[c] (20) (20,-10)
					\Align[c] (40) (40,-10)
					\Align[c] (60) (60,-10)
					\Align[c] (80) (80,-10)
					\Align[l] (\#users) (90,-10)

					\Align[c] (+) (10,13)
					\Align[c] (+) (20,13)
					\Align[c] (+) (30,16)
					\Align[c] (+) (40,17)
					\Align[c] (+) (50,18)
					\Align[c] (+) (60,21)
					\Align[c] (+) (70,26)
					\Align[c] (+) (80,30)

					\Align[r] (100) (-5,10)
				%	\Align[r] (20) (-5,20)
					\Align[r] (300) (-5,30)
					\Align[r] (ms) (-5,50)

				\cip$$
				\vspace{1px}
			
			\end{minipage}
			\caption{Average time computer on function of number of user for 1000 operations}
		\end{subfloat}
		\hspace{5pt}
%		\hspace{1pt}
		\begin{subfloat}
			\begin{minipage}{0.3\linewidth}
				$$\pic

					\xaxis[0,80] 
					\yaxis[0,100] 
					\Line (10,-2) (10,2) 
					\Line (20,-2) (20,2) 
					\Line (30,-2) (30,2) 
					\Line (40,-2) (40,2) 
					\Line (50,-2) (50,2) 
					\Line (60,-2) (60,2) 
					\Line (70,-2) (70,2) 
					\Line (80,-2) (80,2)
					\Line (-2,10) (2,10) 
					\Line (-2,20) (2,20) 
					\Line (-2,30) (2,30) 
					\Line (-2,40) (2,40) 
					\Line (-2,50) (2,50) 
					\Line (-2,60) (2,60) 
					\Line (-2,70) (2,70) 
					\Line (-2,80) (2,80) 
					
					\Align[r] (0) (-5,-10)
					\Align[c] (2k) (20,-10)
					\Align[c] (4k) (40,-10)
					\Align[c] (6k) (60,-10)
					%\Align[c] (8k) (80,-10)
					\Align[l] (\#edges) (70,-10)

				%	\Align[r] (100) (-5,10)
					\Align[r] (200) (-5,20)
					\Align[r] (400) (-5,40)
					\Align[r] (600) (-5,60)
					\Align[r] (800) (-5,80)
					\Align[r] (ms) (-5,90)
									\Align[c] (+) (10,13)
					\Align[c] (+) (20,20)
					\Align[c] (+) (30,36)
					\Align[c] (+) (40,52)
					\Align[c] (+) (50,69)
					\Align[c] (+) (60,89)

				\cip$$
%				\vspace{1px}
			
			\end{minipage}
			\caption{Average time to fill a document for 20 users on all Site}
		\end{subfloat}
		\hspace{5pt}
				\begin{subfloat}
					\begin{minipage}{0.3\linewidth}
						$$\pic

							\xaxis[0,100] 
							\yaxis[0,60] 
							\Line (10,-2) (10,2) 
							\Line (20,-2) (20,2) 
							\Line (30,-2) (30,2) 
							\Line (40,-2) (40,2) 
							\Line (50,-2) (50,2) 
							\Line (60,-2) (60,2) 
							\Line (70,-2) (70,2) 
							\Line (80,-2) (80,2)
							\Line (-2,10) (2,10) 
							\Line (-2,20) (2,20) 
							\Line (-2,30) (2,30) 
							\Line (-2,40) (2,40) 

							\Align[r] (0) (-5,-10)
							\Align[c] (20k) (20,-10)
							\Align[c] (40k) (40,-10)
							\Align[c] (60k) (60,-10)
							\Align[c] (80k) (80,-10)
							\Align[l] (\#ops) (90,-10)

							\Align[c] (+) (10,13)
							\Align[c] (+) (20,12)
							\Align[c] (+) (30,13)
							\Align[c] (+) (40,12)
							\Align[c] (+) (50,11)
							\Align[c] (+) (60,10)
							\Align[c] (+) (70,13)
							\Align[c] (+) (80,12)

							\Align[r] (100) (-5,10)
						%	\Align[r] (20) (-5,20)
							\Align[r] (300) (-5,30)
							\Align[r] (ms) (-5,50)
						\cip$$
						\vspace{1px}

					\end{minipage}
					\caption{Average computing time to execute 10 000 operations function of the total number of operation performed}
				\end{subfloat}
		\caption{Prototype statistic}
		\label{fig:measures}
\end{figure}
The reader can see that execution time is almost linear. Furthermore memory
consumption (not shown here) is directly related to the size of the document (since we use no history file when for GOTO has a quadratic complexity).

{\bf Future works:} We plan to extend this word by adding type information
like DTD or XML schemas which are used to ensure that XML documents comply
with for general structure. The second main extension that we investigate is
the ability to {\em undo} some operations, which may require a limited use of
an history file to recover missing information (needed for instance to recover
a deleted tree).
%>>>>>>> 1.31
%Screenshot
%Stat 
%lien de telechargement
%\section{Discussion \& future work}%{Conclusion}
% autres operation interressante (move)
%Schema/Typage
% Undo
%\section*{toto}
\renewcommand{\refname}{\textsc{\textbf{Reference}}}
\bibliographystyle{acm}
%\singlespacing
\begin{small}
\bibliography{biblio}
\end{small}
\end{document}